\documentclass[journal,sensors,accept,moreauthors,pdftex]{Definitions/mdpi} 

\definecolor{red}{rgb}{0, 0, 0}

\firstpage{1} 
\pubvolume{1}
\issuenum{1}
\articlenumber{0}
\pubyear{2022}
\copyrightyear{2022}
\externaleditor{{Academic Editor: Paolo Bellavista} 
} 
\datereceived{21 January 2022} 
\dateaccepted{11 March 2022} 
\datepublished{} 
\hreflink{https://doi.org/} 
\usepackage{listings}
\usepackage{multirow}
\usepackage[justification=centering]{caption}
\usepackage{makecell}

\usepackage{listings}
\usepackage{comment}

\usepackage{xcolor}

\usepackage{tikz}
\usepackage{pgfplots}\pgfplotsset{compat=1.9}
\usepackage{soul}

\Title{{GLASS}
: A Citizen-Centric Distributed Data-Sharing Model within an E-Governance Architecture}

\TitleCitation{GLASS: A Citizen-Centric Distributed Data-Sharing Model within an E-Governance Architecture}

\Author{ Owen Lo \orcidB{}, William  J. 
 Buchanan{\orcidA{}}, 
Sarwar Sayeed *\orcidC{}, Pavlos Papadopoulos \orcidD{}, \mbox{Nikolaos Pitropakis *\orcidE{}}\mbox{ and Christos Chrysoulas \orcidF{}}}

\AuthorNames{{Owen Lo, William J. Buchanan}, Sarwar Sayeed, Pavlos Papadopoulos, Nikolaos Pitropakis and Christos Chrysoulas}

\AuthorCitation{Lo, O.; Buchanan, W.J.; Sayeed, S.; Papadopoulos, P.; Pitropakis, N.; Chrysoulas, C.}

\address[1]{Blockpass ID Lab
, School of Computing, Edinburgh Napier University, Edinburgh EH10 5DT, UK; {o.lo@napier.ac.uk (O.L.);  b.buchanan@napier.ac.uk (W.J.B.); p.papadopoulos@napier.ac.uk (P.P.); c.chrysoulas@napier.ac.uk (C.C.)} 
}

\corres{\hangafter=1 \hangindent=1.05em \hspace{-0.82em} Correspondence: s.sayeed@napier.ac.uk (S.S.); n.pitropakis@napier.ac.uk (N.P.); {Tel.:} 
 +44-131-455-2789{(N.P.)}}

\abstract{ E-governance is a process that aims to enhance a government's ability to simplify all the  processes that may involve government, citizens, businesses, and so on. The rapid evolution of digital technologies has often created the necessity for the establishment of an e-Governance model.
There is often a need for an inclusive e-governance model with integrated multiactor governance services and  where a single market approach can be adopted. e-Governance often aims to minimise bureaucratic processes, while at the same time including a \emph{{digital}
-by-default} approach to public services. This aims at administrative efficiency and the reduction of bureaucratic processes. It  can also improve government capabilities, and enhances trust and security, which brings confidence in governmental transactions.
However, solid implementations of a distributed data sharing model within an e-governance architecture is far from a reality; hence, citizens of European countries often go through the tedious process of having their confidential information verified. 
This paper focuses on the sinGLe sign-on e-GovernAnce Paradigm based on a distributed file-exchange network for security, transparency, cost-effectiveness and trust 
(GLASS) model, which aims to ensure that a citizen can control their relationship with  governmental agencies. 
The paper thus proposes an approach that integrates a permissioned blockchain with the InterPlanetary File System (IPFS). This  method demonstrates how we may encrypt and store verifiable credentials of the GLASS ecosystem, such as academic awards, ID documents and so on, within IPFS in a secure manner and thus only allow  trusted users to read a blockchain record, and obtain the encryption key. This allows for the decryption of a given verifiable credential that stored on IPFS. This paper outlines the creation of a demonstrator that proves the principles of the GLASS approach.}


\begin{document}



\section{Introduction}  
Our current age is often known as the digital era, where data is a significant resource that scales up constantly. 
Data contains sensitive information that could relate to the confidential information of users, businesses, stakeholders, and so on. 
In a simple data-sharing model, often one entity owns a document and gives access rights to other entities, which could be achieved through role-based access control (RBAC)~\cite{RBAC} or attribute-based access control (ABAC)~\cite{Attribute_Based,Attribute_Based2}.
Data security is a method that aims to protect confidential information from unauthorised access~\cite{GDPR_2}. Data handled by public authorities must be protected when it is associated with individuals' sensitive details.
However, a lack of cyber situational awareness and abuse of sensitive data  can cause  great concern for citizen privacy. Moreover, a third party, such as a public authority and/or agency, often governs the way confidential data is maintained, and due to manual business flows and sluggish interaction methods, a business often fails to build trust and confidence with its partners~\cite{third_party_data, third_party_data2}. Hence,  existing data-sharing approaches often tend to be a serious threat to citizen privacy. 

Achieving the full potential of data sharing can rely on the deploying of  technologies to securely and efficiently collect and transmit the relevant data, thus making them accessible to authorised users~\cite{MehdiSookhak}. This data could be related by an electronic health record (EHR) in an e-Health domain, or via a citizen's confidential identity information. One of the main issues in many existing data-sharing scenarios is that it puts the user's privacy and information security at risk of a data breach. Many organisations have thus adopted centralised systems for data management and sharing, but this can result in various attacks due to a single point-of-failure (SPoF) approach~\cite{centralised_data2,centralised_data}. Although some decentralised distributed methods indicate promising outcomes~\cite{9102377, 7784641, LIU2020102059, 8954646}, the majority of these methods are either research prototypes or based on weak implementations, resulting in \mbox{significant challenges. }

In order to protect sensitive or confidential information, various data protection principles are often in place to require the participating parties to adhere to specific regulations, such as General Data Protection Regulation (GDPR), 
 ({\url{https://gdpr-info.eu/} accessed on 15th February, 2022}~\cite{what_is_GDPR}). Overall, GDPR is endorsed by the European Union (EU) and is a privacy and security law that enforces certain policies to organisations when acquiring user's data~\cite{GDPR_2}. One of the key principles of GDPR is that data cannot be processed if there is no consent from the subject. Regardless of various principles to protect data, a data breach may still occur due to neglecting potential risks or adopting weak methods of securing the data~\cite{Data_risk}.
According to a 2021 data breach report by IBM, the average global cost of a data breach per organisation was USD 4.24 million~\cite{IBM_data_breach}. The report also suggests that the data breach cost in the United Kingdom rose from USD 3.90~million in 2020 to USD 4.67~million in 2021. Generally, centralised approaches to data sharing are one of the main factors that can cause serious threats to citizen privacy. 

In a distributed environment, a collection of computing devices can be connected to share confidential and sensitive data and reduce communication overheads. To overcome this, blockchain can be used as a decentralised, distributed, immutable ledger technology in a trustworthy way~\cite{sayeed2019assessing}. 
Along with this, sectors such as healthcare, voting, asset management, and insurance have shown great improvements in terms of efficiency and security~\cite{stamatellis2020privacy}. Blockchain has also solved other real-world problems, such as in cross-border payments, identity theft, and so on. It can thus also help in protecting confidential data within centralised government systems and within public sectors~\cite{blockchain_solve_prob}. 

A public blockchain is an open network allowing any party to be part of the ecosystem, whereas a permissioned blockchain authorises only the permitted entities to be part of the network. A permissioned blockchain comprises of an access-control method where an administrator grants permissions only to the authorised entities to join a secure channel.




\subsection{e-Governance}
An e-Governance solution primarily focuses on providing a transparent, efficient and coherent service to the citizens~\cite{egovernance,cross_border}. It can connect various entities, such as citizens, government agencies, and trusted institutions into a single domain. The main aim is to enhance the quality of government services so that it improves citizen access to governmental processes through digital methods.  
An e-Governance solution often improves government capabilities, behaviour and professionalism, and such as in the trust and confidence in governmental transactions.

Sharma et al.~\cite{sharma2021unpacking} define that these types of approaches to improved e-governance are \emph{{crucial} 
 for improving the bureaucratic relationship between the government, citizens, and policymakers} and \emph{enforce accountability in governance processes by improving public access to information and transparency}. Krimmer et al.~\cite{krimmer2021developing} define that a key factor of achieving a digital single market requires a strong focus on cross-border integration with the development of e-governance approaches.


\subsection{Research Context}
This research puts forward the distributed scalable and secure e-Governance infrastructure from the scope of GLASS~\cite{DBLP:journals/corr/abs-2109-08566}. GLASS focuses on developing an e-Governance framework that could be adopted  by European Union’s member states. 
It analyses the key technologies of a distributed ledger system that can provide efficient, transparent, cross-border functional, user-friendly techniques to support public services for governments, citizens and businesses across the EU. The paper proposes a security triplet,  which protects the privacy of a stored verifiable  credential. It demonstrates the integration of Hyperledger Fabric with IPFS. While IPFS is used, the design supports the storage of verifiable credentials to any location. The security triplet takes the form of a Content Identifier (CID), a protected encryption key and a URI (Universal Resource Identifier). These are then recorded within a protected Hyperledger Fabric record.



\subsection{Contributions and Paper Layout}
The main contributions of this paper can be summarised as follows:

\begin{itemize}
\item A brief analysis of the key technologies that are essential to shaping a distributed data-sharing model architecture. The paper also provides an overview of the GLASS architectural model to show how various components are integrated. 
\item The creation of an e-governance portal which demonstrates distributed ledger technology into a private instance of an IPFS network. This records the storing of an encrypted verifiable credential to a permissioned blockchain ledger. The actual implementation of this in production will be through a citizen wallet and chaincode.
\end{itemize}


Section~\ref{sec:Background} provides basic concepts about decentralised distributed ledger technology. It also discusses the key components that are required to form a distributed data-sharing model. Section~\ref{sec:Related_Works} reviews the most recent literature indicating the most recent research and gaps. Section~\ref{sec:architecture} outlines a generic architecture of the GLASS model. It also provides a workflow visualising the workings of the proof-of-concept. Section~\ref{sec:proof_of_concept} discusses the architectural development focusing on the security triplet. It also provides performance metrics for the {\it{create}
}, {\it read}, and {\it decrypt} operations. Finally, the paper concludes with Section~\ref{sec:conclusion} discussing the overall contribution of the research work. 


\section{GLASS Background}\label{sec:Background}
\begin{figure}[H]
\includegraphics[width=0.6\linewidth]{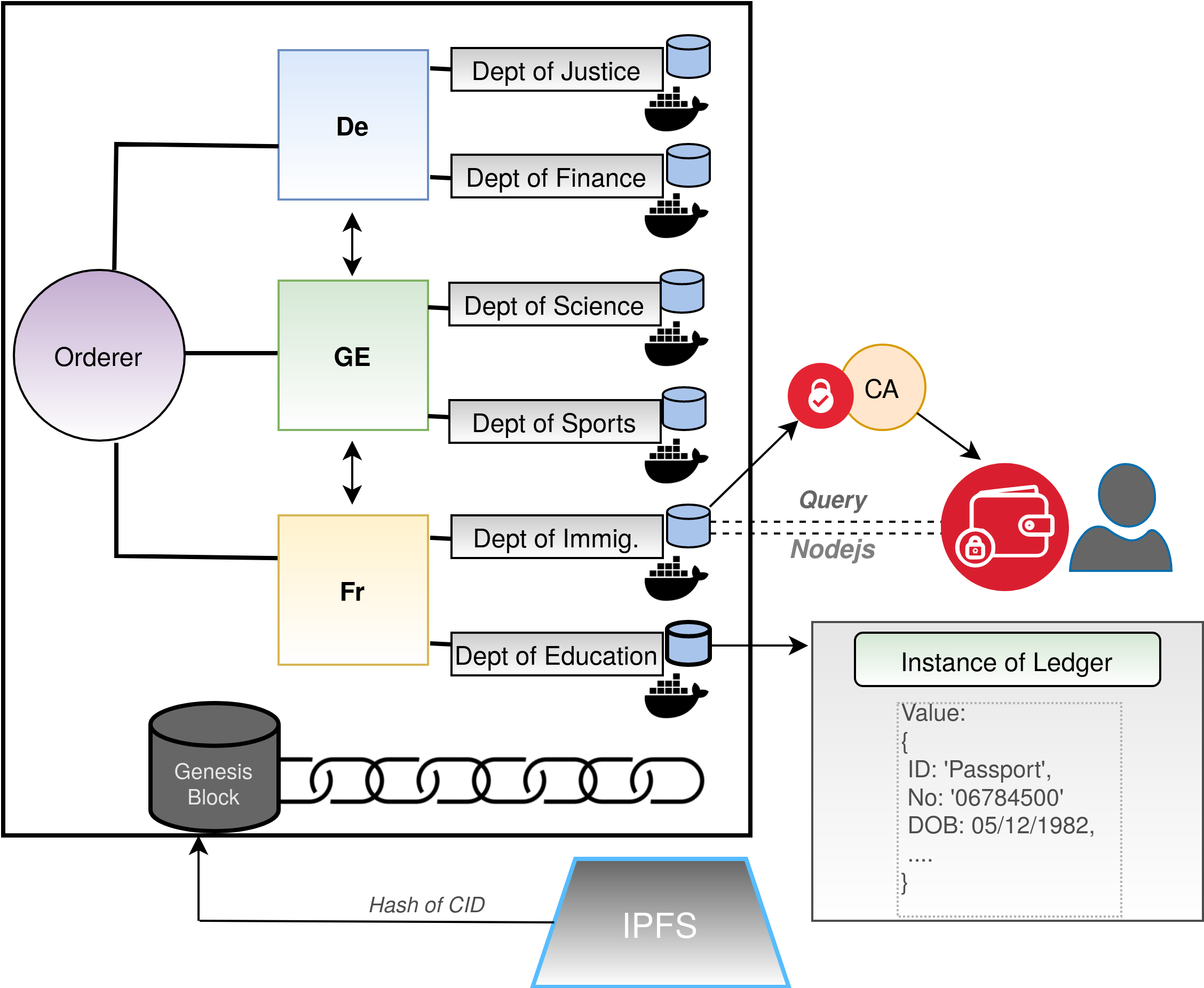}
\caption{An overview of a distributed data-sharing platform  based on permissioned blockchain.}
\label{jain1}
\end{figure}

Figure~\ref{jain1} outlines a generic example of a cross-border data-sharing approach that focuses on the blockchain aspect of an e-governance architecture. This architecture is created using Hyperledger Fabric and  contains three distinct sovereign nations that have joined a single channel to expedite the bureaucratic process. Each sovereign nation comprises two departments that are accountable for endorsing, validating, and committing users'/citizens' data to the blockchain.  A user thus makes a request to the {\it Department of Immigration} using their wallet to retrieve passport information, and also save it in the wallet to prove their identity to another authority. The user’s information can then be stored in the {\it Department of Education's} ledger, and then verified when created.

\subsection{Decentralised Distributed Ledger and Chaincode}
Blockchain is a decentralised, trustless, tamper-proof, distributed ledger over peer-to-peer networks~\cite{ZibinZheng, Papadopoulos2020}. The blockchain and smart contract technology provide a mechanism for developing efficient access control methods to support secure identification, authentication, and user authorization. In addition, the immutable audit trail of blockchain makes it immutable against integrity issues. On the other hand, the smart contract feature facilitates user authentication and authorisation under programmable policies in a fully distributed environment~\cite{MarkoHolbl}. 

In a permissioned blockchain (such as Hyperledger Fabric) the smart contract interconnects with the ledger, and shapes the core segments of the blockchain system~\cite{chaincode}. 
With Hyperledger Fabric, the chaincode is a piece of code that runs on peers within an organisation. It enables the creation and querying of transactions on the shared ledger, and updates the current state of the ledger. This chaincode can be utilised to generate business contracts, asset definitions or oversee decentralised applications. 

\subsection{Distributed Data Sharing}

Research shows that various data-sharing methods were proposed in a distributed environment. Wang et al.~\cite{wang2018data} define a data-sharing method on a dual blockchain approach and where one chain stores the original data (the data blockchain) and another one stores the transactions related to the data (the trading blockchain). This method allows for the transactions performed on the data to be traced. Within this method, the data is broken into $n$ data blocks. Within a transaction, a data owner then signs a random hash on a previous transaction, and provides the data block number, permission level and the public key of the next data user.



\subsection{Interplanetary File System} 

IPFS \cite{benet2014ipfs, ipfs} is a peer-to-peer versioned file system that stores files and offers a data-sharing approach in a distributed manner. 
In general, IPFS is a peer-to-peer protocol that adopts a content-based addressing approach to resource sharing. It follows a similar approach as BitTorrent, where its distributed nature defines how files can be shared across the network. IPFS can be used to generate a permanent and distributed Web-based system (either public or private).



The IPFS process involves generating a cryptographic hash that can be used as the address, as opposed to a URL approach on our existing Web-based accesses. IPFS thus does not follow a centralised approach, rather the peers on the network are able to distribute the data. Moreover, when a peer downloads the content, it becomes a distributor of that content. Digital content such as those related to directory folders, images, and data stores can be represented by IPFS. IPFS breaks down the resources and stores them in a distributed manner. Each block of data is content-addressed using a CID.

\subsection{Content Identifier}
A content identifier, or CID, is a label used to point to resources in IPFS~\cite{Multiformats2020}. It does not indicate where the content is stored (unlike URLs), but it forms a kind of address based on the content itself. CIDs have a fixed length, regardless of the size of their underlying content. They are basically the SHA-256 cryptographic hash of the contents; hence, any minor changes to the data will produce a different CID. An example of a CID for the string-based content of '\emph{{hello} 
 world}' would be:

\begin{quote}
QmT78zSuBmuS4z925WZfrqQ1qHaJ56DQaTfyMUF7F8ff5o
\end{quote}



\subsubsection{Distributed Hash Table}

A Distributed Hash Table (DHT) is a decentralised key-value lookup table~\cite{benet2014ipfs}. It functions as a distributed database that can store and retrieve information based on the keys that the nodes hold~\cite{Distributed_hash_table}. The main feature of DHT is that it maps a CID to the location of content in IPFS. Moreover, nodes are permitted to connect or leave the network and organise themselves to balance and store data 
without having any central entity involved in the network. The DHT algorithm that has been fulfilled by IPFS is referred to as Kademlia~\cite{maymounkov2002kademlia}.




\subsection{Encryption Standards}
Federal Information Processing Standards Publication (FIPS) 140-3 can be utilised as a baseline for encryption~\cite{FIPS}. The following defines the standards used for encryption within the GLASS project:

\begin{itemize}
\item {{Symmetric} 
 key encryption:} This will use 256-bit Advanced Encryption Standard (AES) with Galois/Counter Mode (GCM) and authenticated encryption with associated data (AEAD). The core advantage of using GCM is related to the speed of encryption (as it outputs a stream cipher). Along with this, the AEAD option can be used to integrate the encrypted content with other related data in a verifiable manner.
\item {{Tunnelling:}}This is used when we tunnel through an untrusted network. The encryption used will be 256-bit AES for the tunnel encryption, SHA-256 for the hashing method Elliptic-curve Diffie–Hellman (ECDH) for key exchange, and Elliptic Curve Digital Signature Algorithm (ECDSA) for digital signing. 
\item {{Digital signing:}} With this, we take a key pair of an entity (the private key and the public key). The entity will then sign the document with either ECDSA or  Edwards-curve Digital Signature Algorithm (EdDSA) with its private key. This signature is then checked with the public key. The supported elliptic curve methods will be at least 256-bits.
\item {{Key protection:}} An encryption key will be protected with  RSA using Elliptic Curve Integrated Encryption Scheme (ECIES), where the key is encrypted with the public key of the entity, and revealed with the private key. 
\end{itemize}

\section{Related Literature}\label{sec:Related_Works}

This section reviews some of the most significant works related to distributed data-sharing method.  
The review emphasizes four key aspects: consent models; IPFS and Hyperledger data sharing, and blockchain approaches to e-governance. 

\subsection{Consent Models}
Liang et al.~\cite{liang2017integrating} used Hyperledger Fabric to share consent information. A user could share healthcare information with insurance companies, in order to obtain an insurance quote. On a data-sharing event, an event record is created as a tuple of {recordhash, owner, receiver, time, location, expiry-date, signature}, and submitted to the blockchain network in order to convert health records into trusted transactions. Every action on the health record is then recorded and is thus accountable. 

\subsection{Ipfs and Data Sharing}
Jaiman et al.~\cite{jaiman2020consent} created a blockchain-based data-sharing consent model for health data, and where smart contracts represent a citizen's consent over their health data. It uses two ontologies: Data Use Ontology (DUO) and Automatable Discovery and Access Matrix (ADA-M). These are defined within the Ethereum blockchain. 
Politou et al.~\cite{politou2020delegated} defines the Right to be Forgotten (RtbF) as a requirement of GDPR, and thus for data to be erased.  They implemented an anonymous protocol for delegated content erasure requests within IPFS, and where erasure is only allowed by the original content provider or their delegates.  Unfortunately, enforcing this across an entire IPFS network is not feasible.

An IPFS-based model for electronic health records (EHR) that is  compliant with existing legislation, such as GDPR, was designed  in~\cite{Verdonck2020decentralized}. By presenting some  business processes, the authors showed that EHR access requests can be managed 
with blockchain and smart contract methods. However, the  implementation of the proposed model was not examined in practice.  Furthermore, the deployers of defined smart contracts and the activators  of their transactions were not specified.

\subsection{Hyperledger and Data Sharing}
Stamatellis et al.~\cite{stamatellis2020privacy}, utilised  the usage of private data collection with Hyperledger Fabric and focused on a privacy-preserving healthcare scenario. With this,  entities can store and share their data with privacy controls. To preserve the anonymity and unlinkability of the credentials, the Idemix---a Zero-Knowledge Proof (ZKP) cryptographic protocol suite---was used. Ichikawa et al.~\cite{ichikawa2017tamper} also used Hyperledger Fabric to store healthcare records that were gathered in mobile devices.  Their project is named PREHEALTH~\cite{stamatellis2020privacy}, and does not use the Idemix suite. Overall their system is able to store data in an immutable ledger but without any privacy protection for end-users. It should be noted that to incorporate the private data collection feature, an update of their system is not possible without a complete re-design of their architecture. Likewise, Liang et al.~\cite{liang2017integrating} utilised Hyperledger Fabric to simulate a real-world scenario with different participating entities. On their system, the represented entities include users, wearable devices, healthcare providers, insurance companies, blockchain networks, and cloud databases. However, later versions of Hyperledger Fabric introduced new challenges that their work needs to address, in the case of an architecture re-design and revision into a more recent version that incorporates the private data-collection feature. The authors conclude with helpful metrics of each query's feasibility on their system, although important technical details are missing.

An IPFS/multicloud system using blockchain was proposed to enhance the integrity 
and availability of healthcare records~\cite{Venkatesan2021secure}. The system kept the metadata of health records in a blockchain and encrypted the original 
records in the  cloud/IPFS to guarantee the records'  availability. It also 
provides a control access mechanism, complying with legislation, for the users 
who want to reach patient records. However, there are several limitations in the 
proposed approach: (i) users may have some difficulties for the creation of 
transactions and re-encryption of records; (ii) hospital administrators as actors 
may locally store eHealth records while patients may be unaware of it. Apart from that, information sharing within health care provides strong use cases around citizen permissions and in supporting interoperability. An EHR often contains highly sensitive  healthcare data which are periodically distributed among healthcare providers, pharmacies and patients for clinical diagnosis and treatment~\cite{dubovitskaya2017secure}. 
Furthermore, critical medical information must be regularly updated and shared where proper consent is provided by the patient. Along with this, it requires a strong availability, fast access, and the appropriate encryption on these records~\cite{dubovitskaya2017secure,papadopoulos2021privacy,abramson2020distributed}.


\subsection{Blockchain Approaches to e-Governance}

Geneiatakis et al.~\cite{9102377} assess the feasibility of e-government service to determine 
the transformation of distributed approach by utilising blockchain technology.  They chose a reference test system of an existing cross-border e-government service that is utilised for trading goods across the European Union.  The research outcome reveals how such an aspect can be constructed into a full blockchain-based system. They have leveraged Hyperledger Fabric with 28 nodes, which represents 28 European Union countries and concludes that such distributed approach is efficient in terms of performance and feasibility.

Liu and Li~\cite{LIU2020102059} propose a framework for cross-border e-commerce supply chains. Their framework focuses on blockchain-based models that include a multichain structure model, data-management model, and block structure
model. They have developed various core methods, such as the information-anchoring method, key-distribution method, and so on. The outcome of the analysis evidences their effectiveness against various attacks, such as clone attacks, counterfeit tag attacks, and counterfeit product attacks. The research emphasises both the theoretical and practical aspects of blockchain technology, cross-border e-commerce and supply-chain management.

\section{GLASS Architecture and Methodology}\label{sec:architecture}

This section provides an overview of a distributed data-sharing model on an e-governance architecture. The architecture concentrates on the scope of the GLASS data-sharing model and  gives an overview of the elements associated with a demonstrator. 

\begin{figure}[H]
\includegraphics[width=1.0\linewidth]{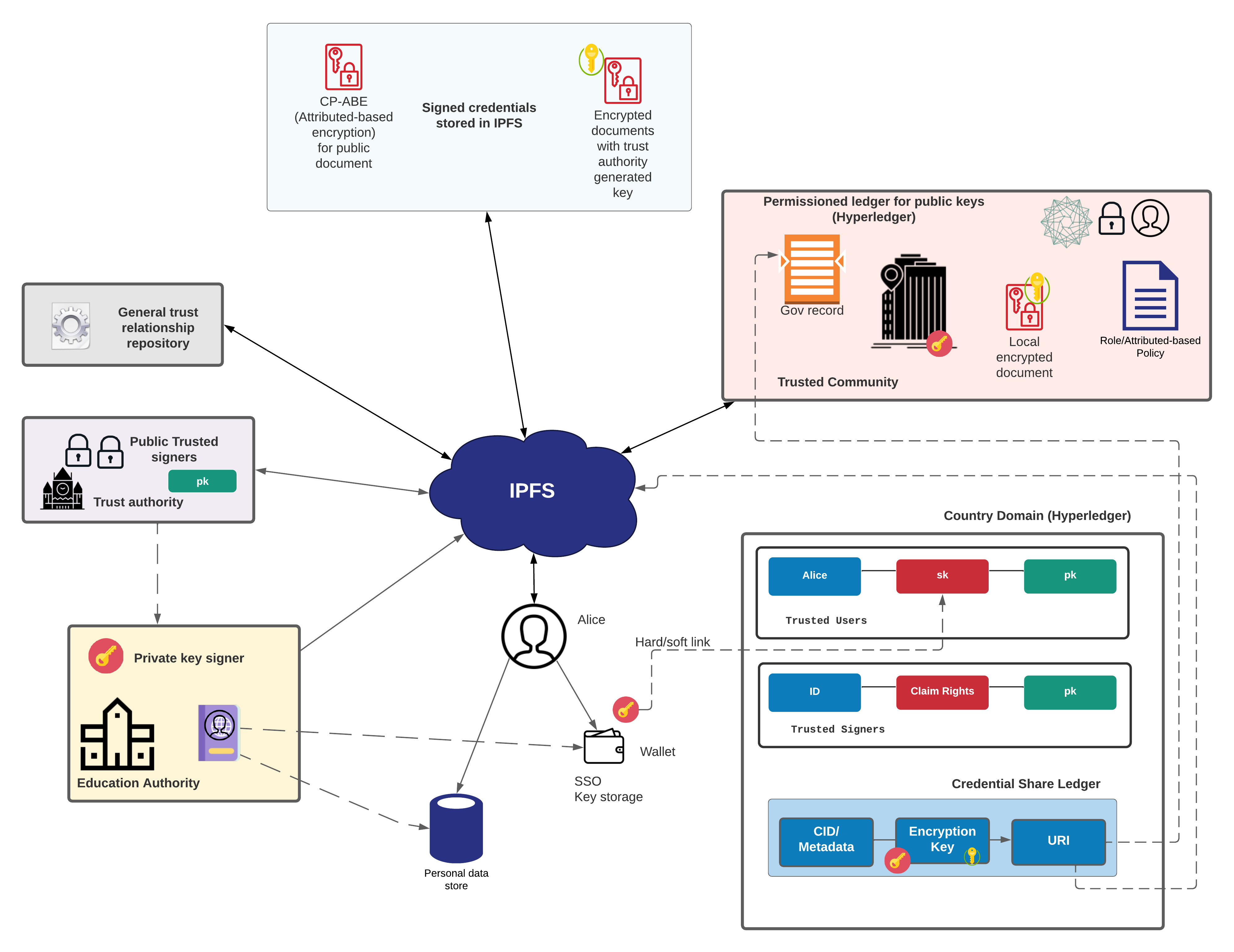}
\caption{Data sharing overview.}
\label{GLASS_overview}
\end{figure}

\textcolor{red}{Figure~\ref{GLASS_overview} provides a generic overview and defines the workings of data sharing within the GLASS architecture~\cite{glass_definition}.}
In this data-sharing model, the verifiable credentials need to be digitally signed by a trusted entity. This will then be passed to the user for them to store them within their wallet or a Personal Data Store (PDS).  
The architecture shows that the Education Authority acts as a trusted authority and passes the credential to the citizen's wallet or PDS. It includes the IPFS protocol, but as IPFS does not support encryption, a tunnel needs to be created between the verifiable credential creator and the target destination (wallet or PDS). Overall verifiable credentials  should not be stored in the IPFS infrastructure without first encrypting them and ensuring an encryption key used is stored outside of the IPFS network---primarily in a permissioned blockchain ledger.


The  GLASS data-sharing model allows citizens to own their verifiable credentials and then provide access rights for another entity. This can be achieved either through role-based access control (RBAC) or attribute-based 
access control (ABAC). With role-based access control, the rights are defined within a domain or across domains based on a role or specific identity. For example, a medical professional could provide access to another medical professional using their security identifier (SID), or could define access through a specific role in the domain. With ABAC, specific attributes can be defined, such as location, time, and role, and then match these to the access policy. To be able to access a document, the user must provide attributes that match the access policy. 
At the core of this, there is a creator of a document, who then defines the owner of the document. The owner has the right to define the access control on the document. This might be controlled within a domain, or can be generally shared.

Attribute-Based Encryption (ABE) is an extension of public-key cryptography which allows one to associate attributes (e.g., name, role, location) with either the cipher-text (known as Key Policy ABE) or private key (known as Cipher Policy ABE). Cipher Policy ABE (CP-ABE)~\cite{bethencourt2007ciphertext} encodes an access policy directly into the encrypted data. Users are  distributed with private keys which have associated attributes (role, name, location). Users who possess the correct attributes and satisfy the cipher policy rule will be able to decrypt the contents of a cipher policy encrypted file. An example of an access policy, for the decryption of a patient's medical record, could be:

\begin{quote}

\end{quote}
\vspace{-0.9cm}
\begin{lstlisting}
(*@\textit{PERMIT IF CHI = 1234567890 OR ROLE = DOCTOR AND LOCATION = UK}@*)
\end{lstlisting}


In a CP-ABE based approach for the GLASS model, all content will be encrypted in the IPFS network and be accompanied by a cipher policy (which dictates who may decrypt the data).  The key advantage to a CP-ABE approach is fine-grain access control on decryption rights. However, certain challenges do need to be addressed including attribute revocation (e.g., how does one revoke a secret key if a doctor is no longer registered?) and the inherent need for trusted authorities (e.g., Government, Hospitals, Legal Departments) to issue private keys embedded with legitimate attributes. A trusted authority might also take on the role of policy execution (such as assigning attributes to private keys and dictating cipher policies) and key management.

CIDs in IPFS are public and are used to address and locate P2P distributed content. Therefore, anyone who knows the CID of a file can retrieve that file by making a download request via IPFS. Although all content hosted on IPFS within the GLASS e-governance model will be encrypted, it is still desirable to limit the number of retrievals to sensitive information since the possibility exists that an attacker may attempt to decrypt the data via malicious means (e.g., brute force attacks, dictionary attacks, social engineer). 
Thus, one approach to reducing this threat is to limit the volume of content that is stored directly in the IPFS protocol. Instead, the CID can be chosen as a pointer to sensitive files which are located in a trusted authority domain (e.g., government portal, hospital servers). The key advantage of this approach is that it reduces exposure of sensitive information and content to the IPFS protocol, which is public by design and immutable given its P2P nature. 

This GLASS model utilises a permissioned blockchain infrastructure to support decentralized identifiers (DIDs), and ABAC. A key element of this is the ability to define private spaces for documents. The Hyperledger Fabric will be used to store the IPFS identifiers. The access control policy can be configured with the Private Data Collection. The topology of the system needs to be designed and the associated technology can be adapted to it. Within the GLASS ecosystem, there are several participating entities such as governmental departments, various organisations and companies, citizens and more. Each entity can be demonstrated through a separate organisation.

Overall the high-level rules involved for GLASS data-sharing infrastructure are:

\begin{itemize}[leftmargin=7.5mm,labelsep=0.5mm,topsep=3pt]
\item No content should be added to the IPFS network unless encrypted, and the encryption key used should be stored outside the IPFS network;
\item IPFS can be used as a tunnel for file transfer between source and destination;
\item Permission ledgers should hold community-related trust keys.
\end{itemize}

\subsection{Workflow Description}\label{sec:workflow}

\subsubsection{Upload, Encrypt, Distribute, and Record Resource}

\begin{figure}[H]
\includegraphics[width=.7\linewidth]{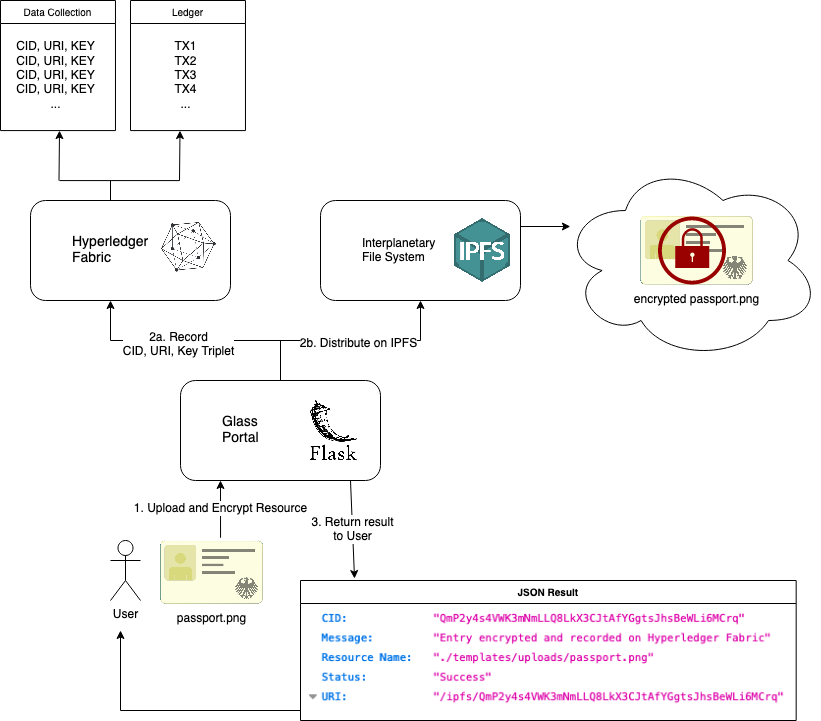}
\caption{An overview of upload, encrypt, distribute and record user resources.}
\label{fig:upload_encrypt}
\end{figure}

Figure~\ref{fig:upload_encrypt} shows a step-by-step visualisation of how the distributed components allow a user to upload, encrypt, and distribute a resource on IPFS; and record the metadata (i.e., the triplet CID, Encryption Key and URI) on Hyperledger Fabric.

\begin{itemize}[leftmargin=7.5mm,labelsep=0.5mm,topsep=3pt]
\item A user uploads their sensitive resource (e.g., a photocopy of their passport) to the GLASS portal. The portal first automatically encrypts the resource using AES-256 CBC. The original resource (plaintext) is deleted from the server once encryption \mbox{is successful.}
    
\item Steps {\it 2a + 2b} occur at the same time. The GLASS portal distributes the encrypted resource on the (private) IPFS network. In doing so, a CID will be generated for the encrypted resource. The CID also acts as a URI since the IPFS uses the CID to both identify (hash) and locate resources. In conjunction with distributing the encrypted resource on IPFS, the GLASS portal also interacts with Hyperledger Fabric to record the triplet of CID, URI and encryption key used to encrypt the resource.
    
\item The status result is returned to the user. If the actions were successful, the user is provided with a CID and URI of their encrypted resource. This allows them or other permissioned individuals to locate it in the future.
\end{itemize}

\subsubsection{Obtain Encryption Key and Decrypt Resource from IPFS}

In order for a user to query Hyperledger Fabric and obtain the encryption key for a resource, the resource can be retrieved from IPFS and decrypted by the GLASS 
portal by inputting the correct key and IV. The steps to obtain the encryption key and decrypt resource from the private IPFS are:




\begin{itemize}
\item The user provides a CID for the GLASS portal to query against Hyperledger Fabric;
    
\item If the user has satisfied the correct policy, 
permission will be granted to allow the GLASS portal to read the encryption key which is mapped to the given CID;
    
\item The encryption key is returned to the user. 
The encryption key and IV are simply separated by a semicolon for ease of interpretation;
    
\item Finally, using the known CID, the user can retrieve the encrypted resource from IPFS. Since the user now has the encryption key, this encrypted resource and key can be provided to the GLASS portal for decryption. If decryption is successful, the user is able to download the decrypted content from the GLASS portal and obtain the original photocopy of their passport.

\end{itemize}

\section{Proof-of-Concept and Evaluation} \label{sec:proof_of_concept}

This section demonstrates the architecture development of the GLASS project. 
It focuses on implementing three components: a GLASS portal, Hyperledger Fabric ledger,  and a private instance of the  IPFS network. 

The Proof-of-concept aims to demonstrate how one may integrate IPFS with Hyperledger Fabric. 
It specifically evidences  the integration of Hyperledger Fabric with IPFS using a user front-end, named as 'GLASS portal'. The workflow implemented demonstrates how we may encrypt and store GLASS resources (e.g., passports, drivers license, medical records, and other citizen-centric resources) on IPFS in a secure manner. All resources stored on IPFS are first encrypted by the GLASS portal. 
A triplet, in the form of CID, Encryption Key and URI are then recorded on Hyperledger Fabric. Only authorised individuals may read the Hyperledger Fabric record, obtain the encryption key and use it to decrypt a given resource stored on IPFS.

\textcolor{red}{{\it Minifabric} (\url{https://labs.hyperledger.org/labs/minifabric.html}  accessed on 15th February, 2022), was 
 used to help develop our proof-of-concept. Minifabric is a software tool which enables rapid deployment of Hyperledger Fabric technology for development and testing purposes using automated scripts. Although it is possible to manually configure and deploy a Hyperledger Fabric instance, we instead chose to use Minifabric in order to alleviate this administrative burden.} To align with the current scope of work, the private IPFS instance has been configured on the same environment as where Hyperledger Fabric and the GLASS portal is running.

\subsection{Permissioned Blockchain}

Hyperledger Fabric is used as our permissioned blockchain for the storage of the triplets (CID, Key and URI) for each resource encrypted and distributed on IPFS. The triplets are stored in a data collection while the ledger itself is used to record read/write transactions (for audit and record-keeping purposes). 

In Hyperledger Fabric, we implement a GLASS-ipfs chaincode to allow users to {\it createGlassResource()} and {\it readGlassResourceKey()}. The former function allows a user to insert a new triplet while the latter function allows a user to read an existing triplet to obtain encryption key of a resource.

Given the sensitive nature of encryption keys, and to demonstrate access policies capabilities in Hyperledger Fabric, two generic organisations have been configured in this workflow: org1.org and org2.org. By default, org1.org has full permission to create new triplets and read existing triplet values. On the other hand, org2.org only has permission to create new triplets and cannot read any encryption key. 
The access policy is defined as shown below:
\newpage


Listing~\ref{application_policies} provides an overview of the access policies in both organisations. It shows that the Hyperledger Fabric environment is configured to use two data {collections: } 
{\it `{collectionGlassResources}`}
 and {\it `collectionGlassResourcesKeys`}. The {\it `collectionGlassResources`} is a public data collection, readable by both organisations, whereas
{\it `collectionGlassResourcesKeys`} is private data collection, readable by org1.org only. The former collection stores the CID and URI of the GLASS resources while the latter collection stores the encryption key. These two data collections, in combination, form the proposed {\it Triplet concept}. 
\begin{lstlisting} [frame=tb, label={application_policies},caption={An overview
 of enforced access policies }, captionpos=b]
[
   {
      ``name'': ``collectionGlassResources'',
      ``policy'': ``OR( 'org1-org.member', 'org2-org.member' )'',
      ``requiredPeerCount'': 0,
      ``maxPeerCount'': 3,
      ``blockToLive'':30,
      ``memberOnlyRead'': true
   },
   {
      ``name'': ``collectionGlassResourcesKeys'',
      ``policy'': ``OR( 'org1-org.member')'',
      ``requiredPeerCount'': 0,
      ``maxPeerCount'': 3,
      ``blockToLive'':30,
      ``memberOnlyRead'': true
   }
  ] 

  
\end{lstlisting}




%

\subsubsection{Distributed Infrastructure of Trust}
\begin{figure}[H]
\includegraphics[width=0.6\linewidth]{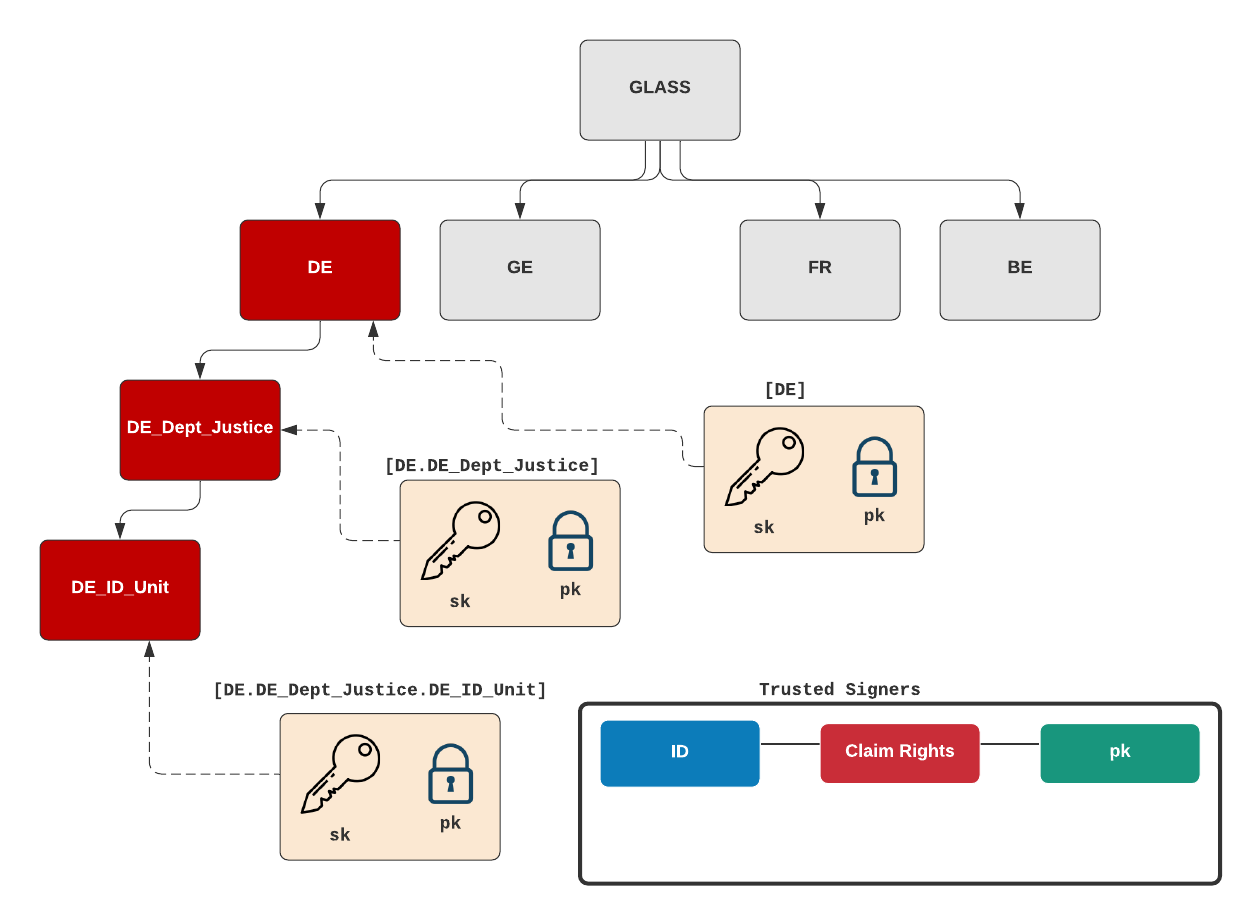}
\caption{Overview of the trust infrastructure.}
\label{fig:trip02}
\end{figure}

A core part of the adoption of the GLASS architecture is in the trust infrastructure based on country-wide domains. Figure~\ref{fig:trip02} outlines this structure in terms of the mapping of country-specific digital signers, and their rights. In the example, we see a top-level domain and then structured into the country domain of [DE]. Trusted organisational units, such as the Department of Justice [DE.DE\_Dept\_Justice] can then map onto this, and where each trusted organisation unit or unit would have their own signing keys. Their associated public key would then be stored within Hyperledger with their common identity (ID), their claim rights.

\subsubsection{Triplets Design Overview}

\begin{figure}[H]

\begin{adjustwidth}{-\extralength}{0cm}
\centering 
\includegraphics[width=1.0\linewidth]{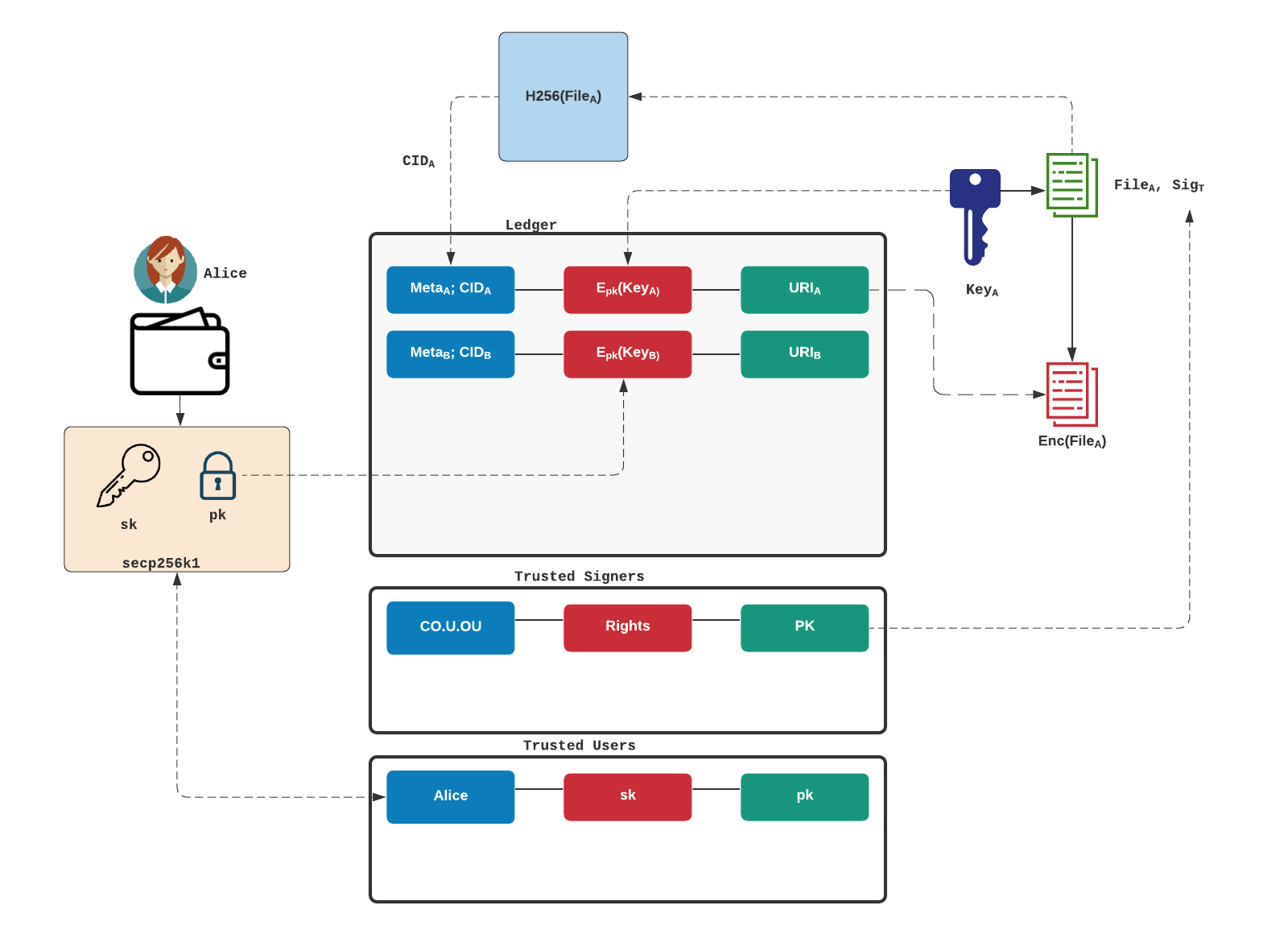}
\end{adjustwidth}
\caption{Overview of triplets.}
\label{fig:trip01}
\end{figure}

The triplets stored on Hyperledger Fabric data collections form the core metadata which allow us to decrypt data distributed on IPFS (or other distribution mechanisms in future work.) 
 Figure~\ref{fig:trip01} outlines the integration of the triplets on the ledger. Each credential file that the citizen (Alice) stores is encrypted with a unique 256-bit encryption key and uses AES GCM. This encryption key is then encrypted with the private key of Alice (and which is either stored in her digital wallet or within the Hyperledger infrastructure). This encrypted key is then stored along with the CID of the credential file and the location of the file (URI---Universal Resource Identifier). This location can either point to an IPFS store, or a URL (Universal Resource Location). 

\begin{figure}[H]
\includegraphics[width=0.95\linewidth]{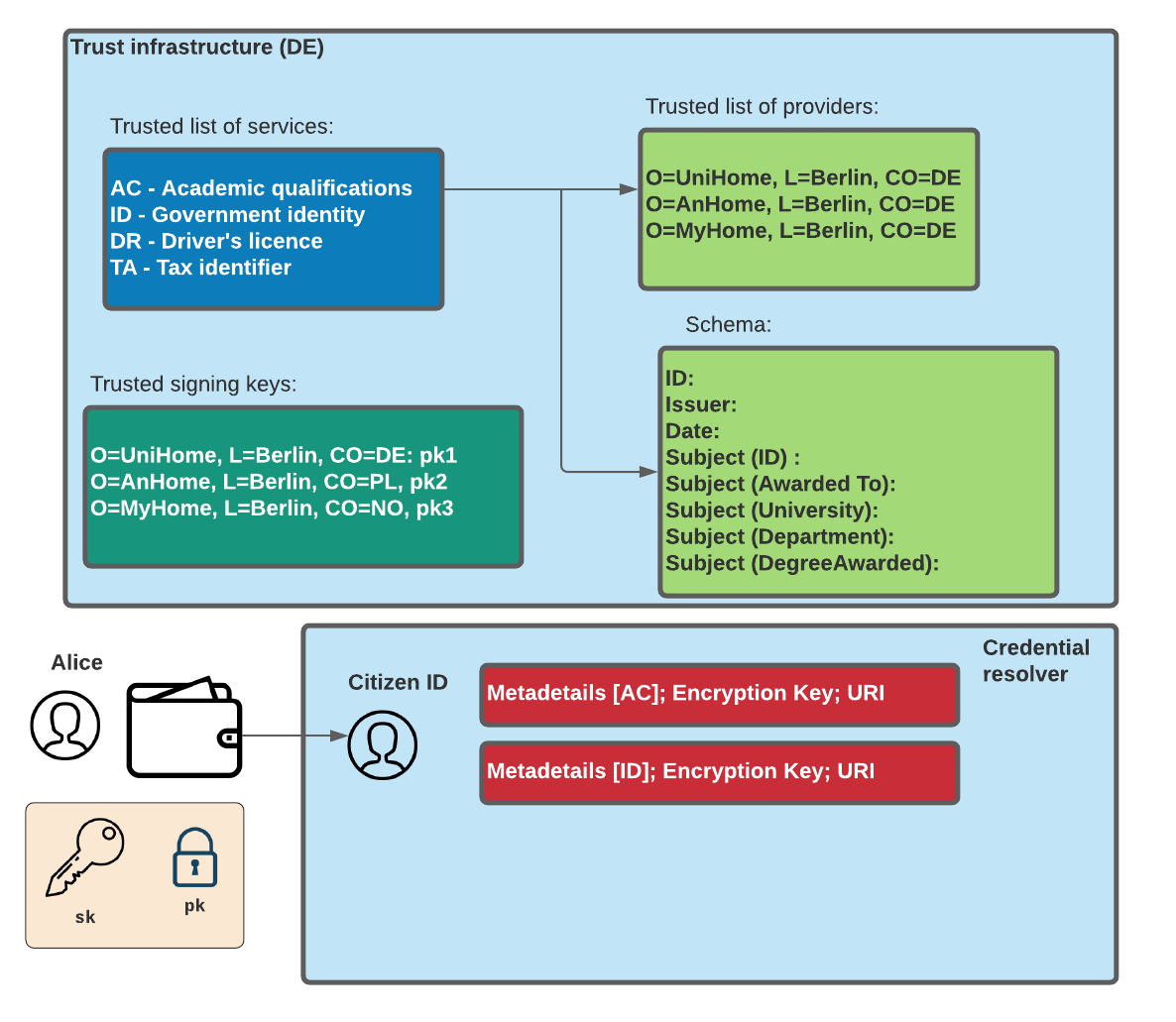}
\caption{Trust infrastructure within ledger.}
\label{fig:trust03}
\end{figure}

\textcolor{red}{In order to support trust within each domain, the public key of the trusted credential signer is stored within each country domain (Figure \ref{fig:trust03}). These are marked as a trusted signers for given credentials, such as AC for Academic Credential. These signers are only trusted for the credential types they have been defined in the trust policy. Each credential is then associated with a credential schema, which is used by the credential signer for core attributes and optional ones. The signer's public key maps to the structure defined in Figure \ref{fig:trip02}. The trust infrastructure focuses on storing the URI for the encrypted credential, but will not have any access to the contents of the file, as the citizen can only rediscover the encryption key using the private key stored in their wallet.}

\vspace{-12pt}

\vspace{-12pt}


Since the focus of this work is the integration of Hyperledger Fabric with IPFS, it may have been noted that the CID and URI will store the same content. As the IPFS protocol uses the CID to both identify and locate resources, this is as expected for IPFS. However, by using the property of URI as a separate field, we may choose to distribute our content in other mechanisms in future work such as Dropbox and Sharepoint (as examples). In such scenarios, the CID of a resource will remain the same but the URI will differ depending on where the resource is located. The encryption key will also remain the same. 
Such architecture allows for an encryption key to be hosted within an external domain.

%
%

%

\subsection{Resource Distribution}

The InterPlanetary File System is a peer-to-peer content-sharing protocol widely used on the internet. This protocol is used as our primary resource distribution mechanism in the developed prototype. All resources distributed on the IPFS network, within our scenario, are encrypted. However, for greater security, a private instance of IPFS was used as the testing ground in the scope of this codebase. A private IPFS functions the same as the public instance of IPFS. However, only nodes that possess a shared private key (referred to as the swarm key) can participate in the private IPFS network. The use of a private IPFS network helps prevent accidental disclosure of confidential or sensitive information.

\subsection{GLASS Portal}

\begin{figure}[H]
\includegraphics[width=.7\linewidth]{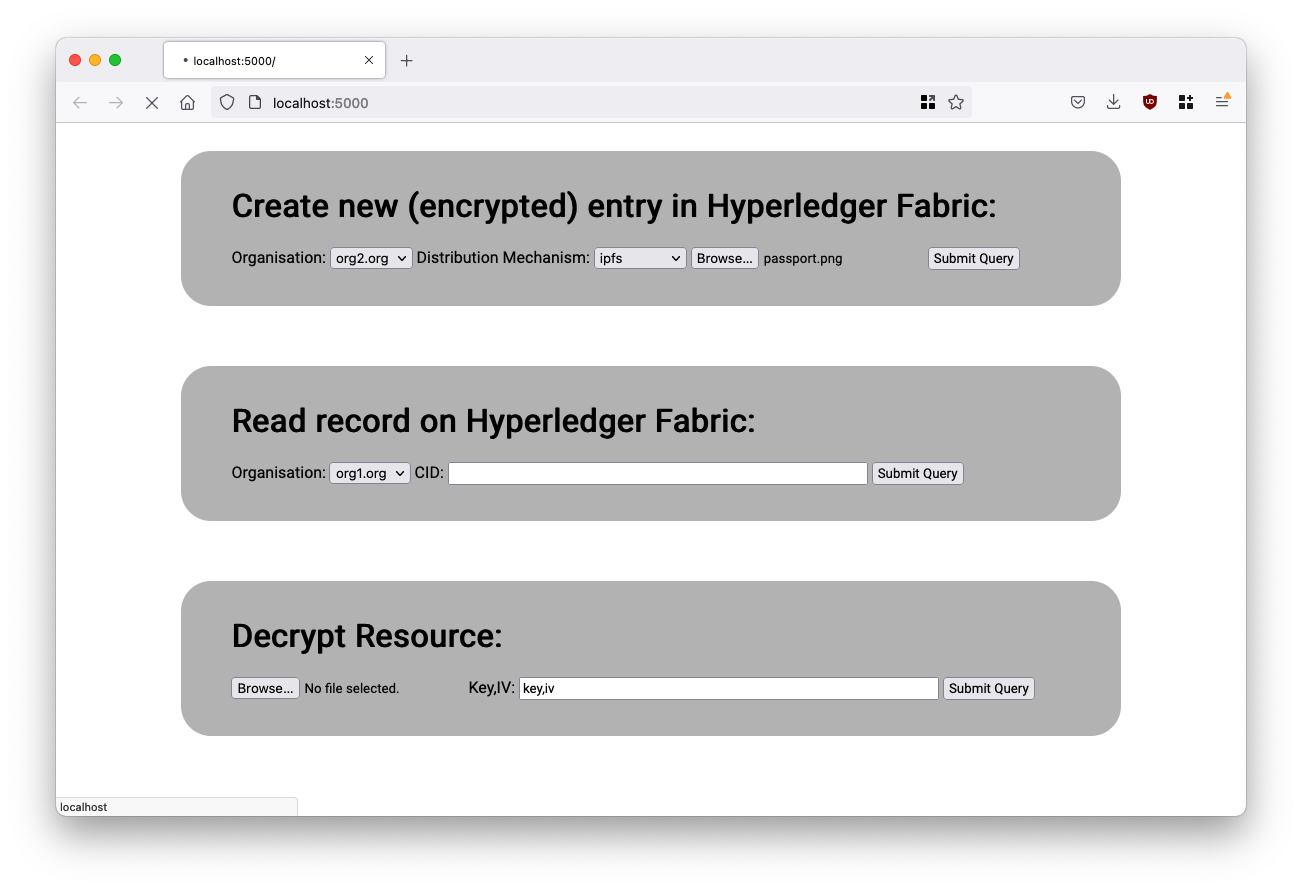}
\caption{Front-end view of GLASS portal associated with three distinct aspects.}
\label{fig:validation}
\end{figure}

The GLASS portal is built on Python Flask (backend) and HTML (frontend). It acts as the interface between Hyperledger Fabric and Private IPFS. The GLASS portal serves as the main component users interact with when uploading GLASS resources (to IPFS) or retrieving GLASS resource metadata (i.e., encryption keys) from Hyperledger Fabric. Figure~\ref{fig:validation} shows the front-end and the three sections which allow the above-described aspects to take place.

A user may use the GLASS portal to safely encrypt and distribute resources on the private IPFS network and query the distributed ledger to obtain encryption keys for resources that are already on IPFS, assuming they have the correct permissions. The three main aspects of the GLASS portal are:

\begin{itemize}[leftmargin=7.5mm,labelsep=0.5mm,topsep=3pt]
\item Allow a user to upload GLASS resources, such as passport, medical data,  to the portal. The portal automatically encrypts and distribute the GLASS resources onto the private IPFS network. AES-256 CBC is used as our encryption function in this proof-of-concept;
    
\item In conjunction with distributing the encrypted resource onto IPFS, the GLASS portal automatically records the triplet metadata (CID, Encryption Key and URI) and stores them in the data collections of Hyperledger Fabric. The permissioned blockchain  records this event in the ledger for audit purposes.;
    
\item \textls[-15]{Read and retrieve encryption keys from Hyperledger Fabric data collection and perform decryption of GLASS resources, assuming the user possesses the correct permissions.}

\end{itemize}

\subsection{Performance Evaluation}

To determine the scalability of our implementation, we measure the time taken to perform the operations to {\it create, read, and decrypt} against an arbitrary resource (229 KB in size). The specification of the machine used to perform this evaluation was a Macbook Pro 2015, 3.1 GHz Dual-Core Intel Core i7, 16 GB RAM and 1 TB SSD.

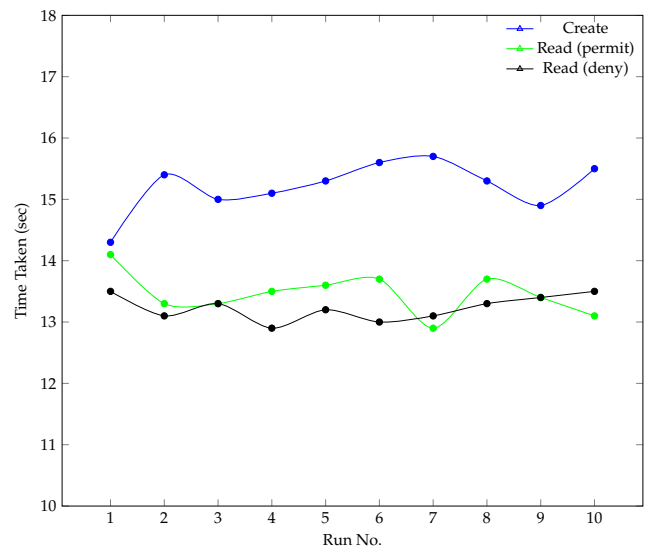
\begin{figure}
\scalebox{0.6}{
\begin{tikzpicture}
  \begin{axis}[ 
  width=\linewidth,
  line width=0.5,
  grid=major, 
  tick label style={font=\normalsize},
  legend style={nodes={scale=1.0, transform shape}},
  label style={font=\normalsize},
  legend image post style={mark=triangle},
  grid style={white},
  xlabel={Run No.},
  ylabel={Time Taken (sec)},
  ytick={10, 11, 12, 13, 14, 15, 16, 17, 18},
  yticklabels={10, 11, 12, 13, 14, 15, 16, 17, 18},
   y tick label style={
    /pgf/number format/.cd,
    fixed,
    fixed zerofill
 },
legend style={at={(1,1)}, anchor=north east,  draw=none, fill=none},
ymin = 10,
ymax = 18,
  ]
    \addplot[blue, mark=*, smooth] coordinates
      {(1,14.3) (2,15.4) (3, 15.0) (4, 15.1) (5, 15.3)(6, 15.6) (7, 15.7) (8, 15.3) (9, 14.9) (10, 15.5)};
      \addlegendentry{Create}
      
    \addplot[green, mark=*, smooth] coordinates
      {(1,14.1) (2,13.3) (3, 13.3) (4, 13.5) (5, 13.6)(6, 13.7) (7, 12.9) (8, 13.7) (9, 13.4) (10, 13.1)};
      \addlegendentry{Read (permit)}
      
   \addplot[red, mark=*, smooth] coordinates
      {(1,13.5) (2,13.1) (3, 13.3) (4, 12.9) (5, 13.2)(6, 13.0) (7, 13.1) (8, 13.3) (9, 13.4) (10, 13.5)};
      \addlegendentry{Read (deny)}
  \end{axis}
\end{tikzpicture}
}
\caption{Time taken to Create and Read.}\label{fig:createRead}
\end{figure}

For each operation, we measured the time taken for 10 independent runs. Figure~\ref{fig:createRead} shows the time taken for each independent run (decrypt times are omitted due to scale), while the total average times calculated for each operation is presented in Table~\ref{tab:avgResults}. In all three operations, we chose to omit the time taken to perform network-related activities (file upload or download) as this variable is outside the control of our experimentation. Additionally, it should be noted that the file size chosen is arbitrary but fixed in size to ensure consistency between each iteration of our experiment. In the context of this work, the operations of {\it Create}, {\it Read}, and {\it Decrypt} are defined as follows:

\begin{enumerate}
    \item Create: in this operation, the total time taken to perform resource encryption, generate and distribute the resource on a private IPFS network and record an entry of the `Triplet` value on Hyperledger Fabric is measured;
    
    \item Read: records the total time taken to read a `Triplet` value from Hyperledger Fabric. The scenario of both permit and deny are measured (i.e., the request is granted and denied respectively based on our access policy defined in the Hyperledger \mbox{Fabric chaincode); }
    
    \item Decrypt: measures the total time taken to decrypt a resource.
    
\end{enumerate}

\begin{table}[H]
\caption{Average time (sec) taken to Create, Read, and Decrypt.}\label{tab:avgResults}
\setlength{\cellWidtha}{\textwidth/2-2\tabcolsep-0.0in}
\setlength{\cellWidthb}{\textwidth/2-2\tabcolsep-0.0in}
\scalebox{1}[1]{\begin{tabularx}{\textwidth}{>{\centering\arraybackslash}m{\cellWidtha}>{\centering\arraybackslash}m{\cellWidthb}}

\toprule

\textbf{Operation}     & \multicolumn{1}{l|}{\textbf{Avg. Time Taken ({S}
)}} \\ 
\midrule
\textit{{Create} 
}        & 15.2                                               \\ 
\textit{Read (permit)} & 13.5                                               \\ 
\textit{Read (deny)}   & 13.2                                               \\ 
\textit{Decrypt}       & 0.002                                              \\ 
\bottomrule
\end{tabularx}}

\end{table}

\vspace{0.7 em}

The results demonstrate very consistent create and read times with a low degree of variability between each instance of run. On average, it takes 15.2 s to encrypt a resource, distribute it on the IPFS network and record the entry (triplet value) on Hyperledger Fabric. Similar times are observed for the operation of reading entries in Hyperledger Fabric with average times of 13.5 and 13.2 s when the operation is permitted and denied respectively. Lastly, decryption times are exceptionally fast at 0.002 s on average. Based on the evaluation conducted, our findings demonstrate that the  majority of processing time is spent creating or reading entries from Hyperledger Fabric. CID generation time, using SHA-256 hashing, is minimal as shown in prior research where up to 100 CIDs were generated in less than 1 s \cite{Barati2021}. However, it is acknowledged that the time taken may be file-size-dependent. Additionally, the processing time for  encryption/decryption functions is also minimal given the small file size. Although it is acknowledged that larger file sizes will increase the time to encrypt/decrypt a file using AES-256 CBC, our implementation is not constrained to using this particular algorithm.  

\subsection{Limitations}

\textcolor{red}{Although the performance evaluation results provide evidence that our PoC  functions within acceptable time frames for {\it Create, Read, and Decrypt} operations, we must acknowledge three main limitations to this current work. Firstly, our performance evaluation has been conducted on a single test machine whereby all three components (GLASS portal, Hyperledger Fabric and Private IPFS) are running. The performance overheads may be introduced in such an experimental setup due to the need for each component to share resources on a single machine. Secondly, the performance evaluation we have conducted is limited to the metric of average time taken for certain operations. Third, only a small sample of test runs was conducted for {\it Create, Read, and Decrypt operations.}}

\section{Conclusions} \label{sec:conclusion}

A cross-border digital single market can consolidate European countries to enhance greater access to data, and initiate substantial feasibility. One of the main objectives of  GLASS is to achieve a citizen-centric distributed data-sharing model which can minimise  bureaucratic processes and build trust among its citizens. 
In this research, we have proposed an approach to integrate Hyperledger Fabric with IPFS. We have also shown the method of encrypting and storing verifiable credentials within IPFS. Besides that, it is shown that only the trusted users with correct permissions have access to blockchain records and can acquire encryption keys.  


For future work, we aim to evaluate encryption functions in further detail including assessing the trade-off between security and performance for different algorithms. We also aim to integrate the three key components with a user wallet to uncover the access control functionalities. \textcolor{red}{We also intend to deploy each component of our work on separate distributed machines to reduce performance overheads alongside monitoring and acquiring a larger sample of evaluation metrics (e.g., network throughput, CPU usage, memory usage of each independent machine) for greater oversight on the scalability of our implementation.}

\authorcontributions {All authors contributed in the conceptualization and methodology of the manuscript; O.L. performed data preparation and the practical experiments; O.L., W.J.B. and S.S., contributed in writing; P.P., C.C. and N.P. reviewed and edited the manuscript. All authors have read and agreed to the published version of the manuscript} 



\funding{The research leading to these results has been partially funded by the European Union's Horizon 2020 research and innovation programme, through funding of the GLASS project (under grant agreement No 959879).}

\conflictsofinterest{The authors declare no conflicts of interest.}

\begin{adjustwidth}{-\extralength}{0cm}
\reftitle{References}



\end{adjustwidth}

\end{document}